\documentstyle[epsfig]{aipproc}

\begin{document}
\title{Two-phase spectral modelling of 1E1740.7-2942 }

\author{Osmi Vilhu$^1$,  Jukka Nevalainen$^1$, Juri Poutanen$^2$, Marat
Gilfanov$^3$, Philippe Durouchoux$^4$, Marielle Vargas$^4$, 
Ramesh Narayan$^5$  and Ann Esin$^5$  } 
\address{$^1$ Observatory, FIN-00014 University of Helsinki, Finland\\
$^2$ Uppsala Observatory, Uppsala, Sweden\\
$^3$ Space Research Institute, 117810 Moscow, Russia\\
$^4$ C.E.Saclay, 91191 Gif-sur Yvette, France\\
$^5$ Harvard-Smithsonian Center for Astrophysics, Cambridge, MA 02138, USA    
                                   }

\maketitle

\begin{abstract}

Combined ASCA and SIGMA data of 1E1740.7-2942 during its standard state
(September 1993 and 1994) were fitted
with two-phase models (ISMBB \cite{Poutanen96,Poutanen97} and ADAF
\cite{Narayan96,Esin97}). 
In ISMBB's, the radius  of the spherical hot (T$_e$ =
150 - 200 KeV)
corona lies between 200 - 250 km where it joins the classical inner disc. 
The disc
 radiates 40 $\%$
of the total luminosity (with $\dot M$ =  0.017$\dot M_{Edd}$ 
of 10M$_{\odot}$). ADAF's need an extra component to reproduce 
the soft part of spectrum.
However, the origin of the soft excess remains somewhat uncertain,
although special care was taken in the background elimination.

\end{abstract}

\section*{Introduction}
1E1740.7-2942 is the famous jet-source (micro-quasar, Great Annihilator) 
close to the Galactic Center,
observable in the cm/mm and X/Gamma-ray wavelengths 
\cite{Sunyaev91,Bally91,Mirabel91,Mirabel92,Sheth96,Churazov96,Vilhu97}. 
During the last few years, new tools for 
modelling of X-ray data have been developed. These include  two-phase
sombrero models ISMBB's  \cite{Poutanen96,Poutanen97} 
and 
ADAF's (advection dominated accretion flows \cite{Narayan96,Esin97}). 
 While ADAF's are fixed 
basically by $\dot M$ (mass transfer rate),  ISMBB's 
 have more freedom  to fit  observations.

\begin{figure}[b!] 
\centerline{\epsfig{file=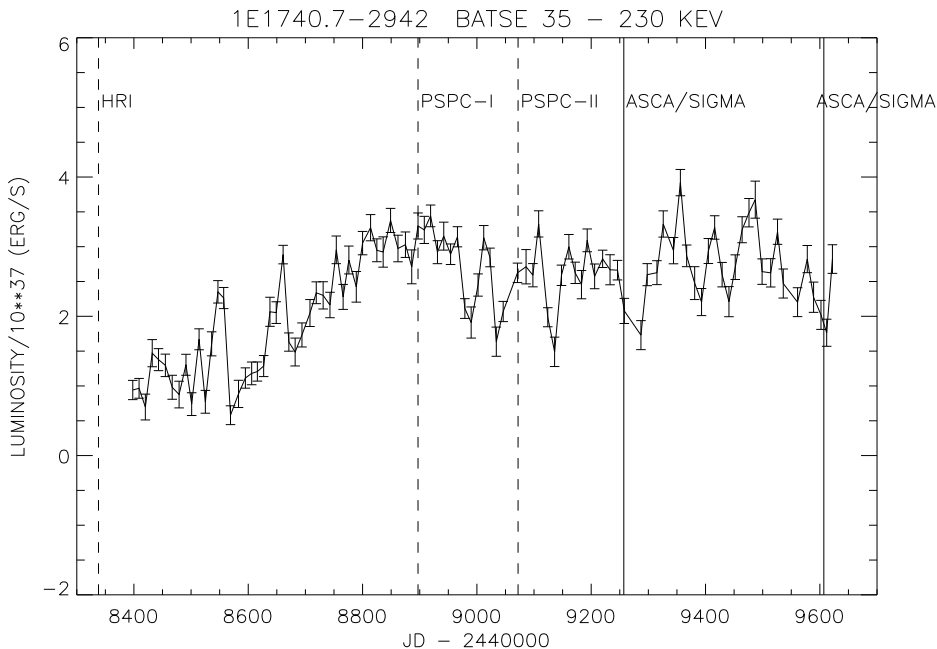,height=3.5in,width=5.0in}}
\vspace{10pt}
\caption{BATSE light curve of 1E1740.7-2942 showing the dates of the 
observations discussed in the text.}
\label{fig1}
\end{figure}

We used ASCA archive data  and  
SIGMA hard X-ray observations (Sept 1993 and 1994)  to explore 1E1740.7-2942 with the help
of these models (see Figure \ref{fig1}). 
Simultaneous BATSE results 
were used for comparison. The near-by background in ASCA images was 
carefully analysed and subtracted.

\section*{ISMBB and ADAF models}

\begin{figure}[b!] 
\centerline{\epsfig{file=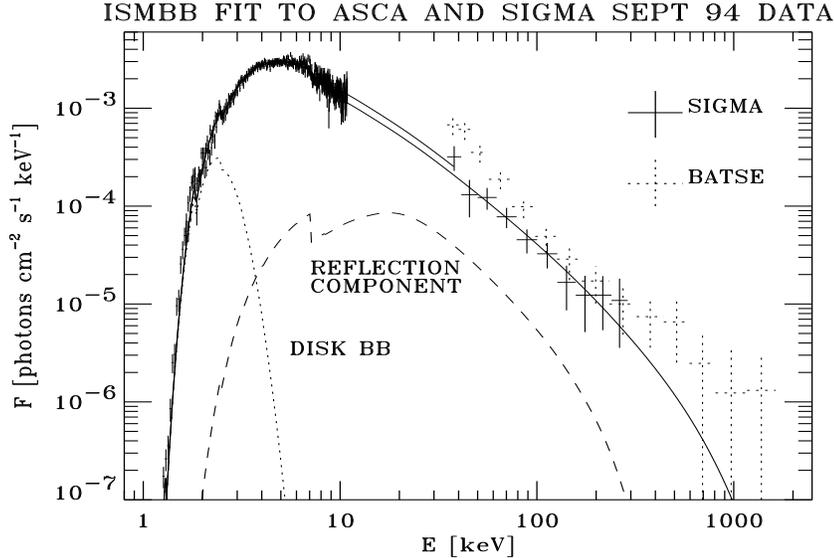,height=3.5in,width=6.0in}}
\vspace{10pt}
\caption{ISMBB fit to the September 1994 data. BATSE results are overplotted. }
\label{fig2}
\end{figure}

 The hot phase of ISMBB
was handled as a pure spherical thermal pair plasma 
(defined by T$_e$ and $\tau$$_e$),
 although the background
plasma (protons) and non-thermal electrons can be included in the treatment.  
To solve 
self-consistently the pair balance
, the energy balance  and the
radiative transfer equations, the 
iterative scattering
method (ISM) was used (\cite{Poutanen96}). 
The source of soft photons is the classical disc
black body radiation  with radial 
temperature-dependence T(r) = T$_{bb}$(r/R$_{in}$)$^{-3/4}$. 
The disc is allowed to reach inside R$_{in}$ 
(where T is assumed constant), but all our fits converged to the case where
the radius of the hot phase equals to R$_{in}$. 
The radiatively heated 
reprocessed radiation is
included in the disc black body, increasing the disc luminosity typically
by 20 $\%$. 
The solution was  found running 
the models and the input data under XSPEC of the XANADU software 
(see Figure  \ref{fig2}  and  Table \ref{table1}).

\begin{table}
\caption{ISMBB and ADAF  fits to the standard state of 1E1740.7-2942 }
\label{table1}
\begin{tabular}{lrrrr}
PARAM& ISMBB 94 & ISMBB 93 & 
ADAF-1 94 & ADAF-1 93 \\
\tableline
N$_H$ (10$^{22}$ cm$^{-2}$) & 12.3 & 12.3 & 15.7          &        16.8 \\
kT$_{bb}$ (KeV)                & 0.25 & 0.23 & $\approx$ 0.2 & $\approx$ 0.2 \\
T$_e$ (KeV)               & 200  & 150  & $\approx$ 110 & $\approx$ 110 \\
$\tau$$_e$                & 0.7  & 1.3  & $\approx$ 2.1 & $\approx$ 2.1 \\
cos(incl)                 &  0.95    & 0.95     &    0.87           &   0.87 \\
R$_{in}$\tablenote{R$_{in}$ = Rc = R$_{TR}$, in R$_{Sch}$ units of 
               10M$_{\odot}$  } & 7 & 8 & 10 & 10 \\
L$_{total}$ (10$^{37}$ erg/s) & 4.5 & 5.6 & 3.5 & 4.4 \\
L$_{disc}$/L$_h$ & 0.82 & 0.55 &  &    \\
red. $\chi$$^2$  & 1.13 & 1.12 & 1.17 & 1.16 \\
\end{tabular}
\end{table}

Several ADAF models were computed around  the intermediate state
 \cite{Esin97} 
with $\alpha$ = 0.3 (viscosity parameter), $\beta$ = 0.5 (magnetic pressure
parameter), m = 10 (mass in solar units) and $\dot m$ = 0.11 (mass transfer
rate in Eddington units with efficiency 0.1, $\approx$ m$_{crit}$). 
These  models are too luminous
by a factor of 4 which can be
easily accounted for by slightly 
reducing the parameters (the 
luminosity is proportional to $\alpha$$^2$m$\dot m$).  
The $\dot m$-value from ISMBB-fits is also smaller.

\begin{figure}[b!] 
\centerline{\epsfig{file=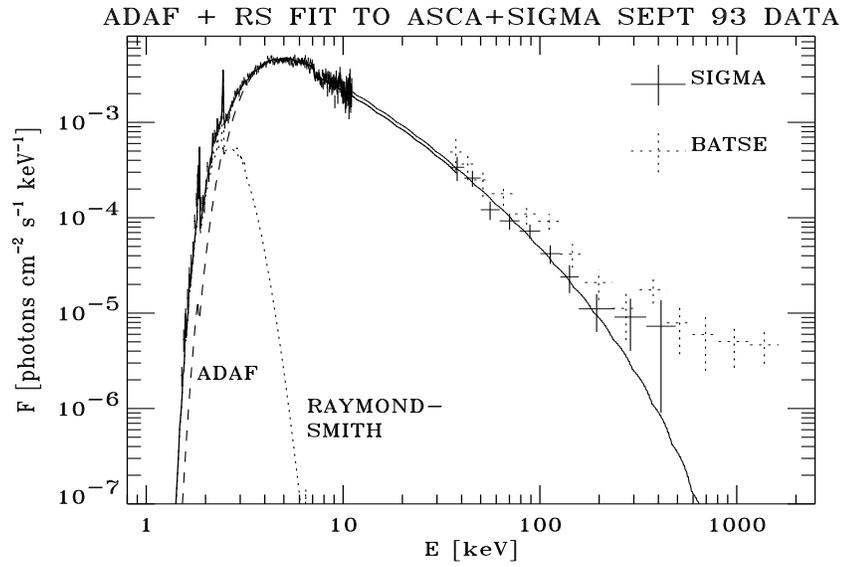,height=3.5in,width=6.0in}}
\vspace{10pt}
\caption{ ADAF model (R$_{TR}$ = 10R$_{Sch}$) fit to the September 1993 data. BATSE results are  overplotted.}
\label{fig3}
\end{figure}

\begin{figure}[b!] 
\centerline{\epsfig{file=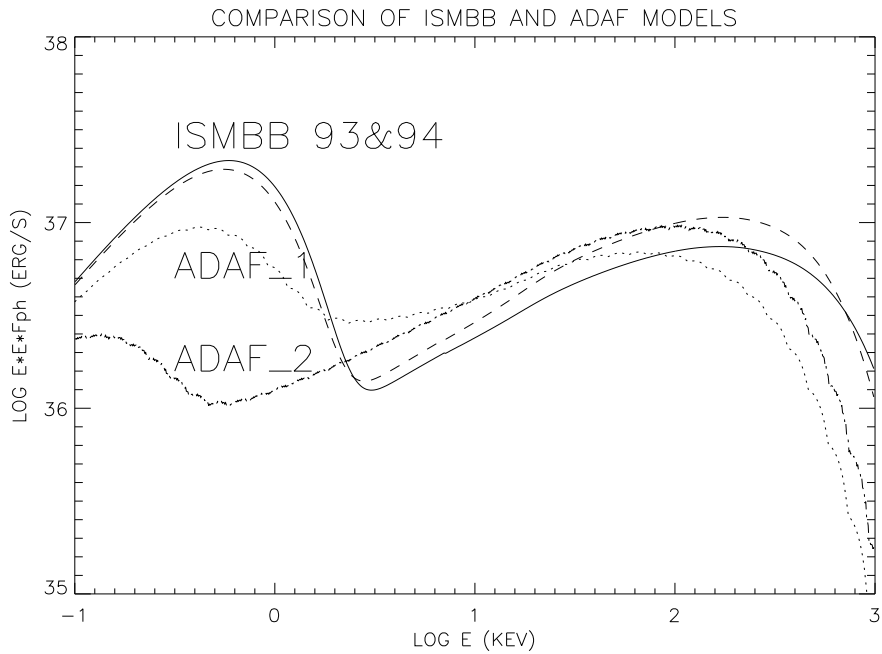,height=3.5in,width=5.0in}}
\vspace{10pt}
\caption{ ISMBB (1993 and 1994)  and  ADAF (R$_{TR}$ = 10 and 100)
                  models.  }
\label{fig4}
\end{figure}

Figure
\ref{fig3} shows the fit  with a model having 
transition radius R$_{TR}$ = R$_{in}$ = 10 (in 
Schwarzchild units). For a smaller radius the spectrum is too soft, while
for a larger one the disc spectrum lies totally 
outside the ASCA-range (see
Figure \ref{fig4} where the models are compared).
Even the model with R$_{TR}$ = 10 needs an extra component,
in Figure \ref{fig3} the excess was modelled by a thin plasma 
(Raymond-Smith) with  emission measure of 1.1$\times$10$^{62}$ cm$^{-3}$
and T = 0.28 KeV. Using the Sedov-model, we can speculate that
a shocked young ($\le$ 1000 years) supernova remnant,
expanding in a dense ($\ge$10$^{2.5}$ cm$^{-3}$) medium and originated in a
relatively quiet explosion (E$_T$ $\le$ 10$^{48.5}$ erg), can explain these
values and the small size (point source in the ROSAT HRI-image).   
   
\section*{Conclusions}
The standard state spectrum of 1E1740.7-2942  can be fitted with the
ISMBB models. The disc has
an inner radius of 7 - 8 Schwarzchild radii of 10M$_{\odot}$ 
black hole
radiating with L$_{disc}$ = 2.2$\times$10$^{37}$ erg/s, corresponding to 
the mass transfer rate of
2.3$\times$10$^{17}$ g/s = 0.017$\dot M_{Edd}$  ($\dot M_{Edd}$ = 
10L$_{Edd}$/c$^2$ = 1.39$\times$10$^{18}$M/M$_{\odot}$ g/s, using 
efficiency of 0.1).
Inside  R$_{in}$, the spherical hot corona (T$_e$ = 150 - 200 keV) 
emits Comptonized
radiation with  
L$_h$ = 2.7$\times$10$^{37}$ erg/s. 

The ADAF models can reproduce the overall spectral shape and luminosity,
 provided that the total
mass, the mass transfer rate and the viscosity parameters 
are properly selected.
For small values of R$_{in}$, the ADAF hot flows are
somewhat too cold and the disc spectrum too shallow (see Figure \ref{fig4}). 
However, it is
possible that the soft excess has some other origin we missed (a local
excess in the background, circumstellar scattering dust, young SNR in a dense
medium, ...), but 
this can be decided only with future 
observations having a higher spatial resolution (like AXAF). 

An important test for the nature of the soft excess may be the observed 
ROSAT count rates
\cite{Heindl94}. The ISMBB models
of Table 1 predict 2.5 times  higher and 0.8 times lower PSPC and HRI count
rates, respectively. If the disc is removed, the predicted count rates are 
reduced by a factor of 3. However,  
if an anticorrelation between the soft and hard luminosities 
exists, a part of the difference can be due 
to a real variability (see Figure \ref{fig1}).


\begin{references}

\bibitem{Bally91}Bally J. and Leventhal M. 1991, Nature 353, 234.

\bibitem{Churazov96}Churazov E., Gilfanov M. and Sunyaev R. 1996, 
                           ApJ 464, L71.

\bibitem{Esin97}Esin A.A, McClintock J.E. and Narayan R. 1997, 
                        astro-ph/9705237.

\bibitem{Heindl94}Heindl W.A., Prince T.A. and Grunsfeld J.M. 1994,
                                ApJ 430, 829.

\bibitem{Mirabel91}Mirabel I.F. et al. 1991, Astron.Astrophys. 251, L43.

\bibitem{Mirabel92}Mirabel I.F. et al. 1992, Nature 358, 215.

\bibitem{Narayan96}Narayan R. 1996, ApJ 462, 136.

\bibitem{Poutanen96}Poutanen J. and Svensson R. 1996, ApJ 470, 249.

\bibitem{Poutanen97}Poutanen J. 1997, private comm.

\bibitem{Sheth96}Sheth S., Liang E., Luo C., 1996. ApJ 468, 755.

\bibitem{Sunyaev91}Sunyaev R. et al. 1991, ApJ 383, L49. 

\bibitem{Vilhu97}Vilhu O. et al. 1997, Proc. 2nd INTEGRAL workshop 'The Transparent Universe',
                         ESA SP-382, p.221, astro-ph 9612194. 

\end{references}
\end{document}